\documentclass[pra,aps,amsmath,amssymb,showkeys]{revtex4}

\newcommand{\hh}{{\mathcal{H}}}
\newcommand{\lnp}{{\mathcal{L}}}
\newcommand{\lsa}{{\mathcal{L}}_{s.a.}}
\newcommand{\lsp}{{\mathcal{L}}_{+}}
\newcommand{\mcn}{\mathcal{N}}
\newcommand{\pen}{\openone}
\newcommand{\id}{{\mathrm{id}}}
\newcommand{\xb}{{\mathsf{X}}}
\newcommand{\yb}{{\mathsf{Y}}}
\newcommand{\kb}{{\mathsf{K}}}
\newcommand{\ab}{{\mathsf{A}}}

\newcommand{\nm}{{\mathsf{N}}}
\newcommand{\bab}{{\mathsf{B}}}
\newcommand{\cab}{{\mathsf{C}}}
\newcommand{\bro}{{\boldsymbol{\rho}}}
\newcommand{\vrbro}{{\boldsymbol{\varrho}}}
\newcommand{\bomg}{{\boldsymbol{\omega}}}
\newcommand{\vm}{{\mathsf{V}}}
\newcommand{\hm}{{\mathsf{H}}}
\newcommand{\veps}{\varepsilon}
\newcommand{\Tr}{\mathrm{Tr}}

\begin{document}
\clearpage
\preprint{}

\title{Non-equilibrium equalities with unital quantum channels}

\author{Alexey E Rastegin}

\affiliation{Department of Theoretical Physics, Irkutsk State University,
Gagarin Boulevard 20, Irkutsk 664003, Russia}

\begin{abstract}
A general tool for description of open quantum systems is given by
the formalism of quantum operations. Most important of them are
trace-preserving maps also known as quantum channels. We discuss
those conditions on quantum channels under which the Jarzynski
equality and related fluctuation theorems hold. It is essential
that the representing quantum channel be unital. Under the
mentioned condition, we first derive the corresponding Jarzynski
equality. For bistochastic map and its adjoint, we further
formulate a theorem of Tasaki--Crooks type. In the context of
unital channels, some notes on heat transfer between two quantum
systems are given. We also consider the case of a finite system
operated by an external agent with a feedback control. When unital
channels are applied at the first stage and, for a
mutual-information form, at the further ones, we obtain quantum
Jarzynski--Sagawa--Ueda relations. These are extension of the
previously given results to unital quantum operations.
\end{abstract}

\keywords{exact results}

\maketitle

\pagenumbering{arabic}
\setcounter{page}{1}

\section{Introduction}\label{sec1}

Complex systems away from thermal equilibrium are one of the
principal issues of statistical physics. Growing interest to
thermodynamic of systems at the nanoscale has also stimulated a
more detailed study of statistical fluctuations \cite{jareq11}. A
particular phenomenon, the so-called ''return to equilibrium'',
was rigorously analyzed \cite{bfs00,fm04}. In \cite{asf05},
adiabatic theorems and their connection with the zero law of
thermodynamics are reviewed. Further, the role of computer
simulations in analyzing non-equilibrium processes will certainly
increase. Many efficient schemes for simulation of complicated
biomolecules and colloidal particles were developed \cite{senn07}.
Due to Jarzynski \cite{jareq97a,jareq97b}, novel advances have
been achieved for thermodynamic systems driven out from the
equilibrium by external forces \cite{jareq11,talhag09}. If these
forces are varied in line with a specified protocol, then some
exact relations can be derived. The first of exact non-equilibrium
relations is now referred to as the Jarzynski equality. Results of
such a kind are significant in own rights as well as for extending
the scope of computer simulations \cite{psh03}.

Let us consider a thermally insulated system, which is acted upon
by a time-dependent external field. For any quasi-static process,
the work performed on the system is equal to the difference
between the final and initial free energies. For a non-equilibrium
process, averaged total work will exceed this difference due to
the second law of thermodynamics. Instead of inequalities,
Jarzynski gave the equality connecting non-equilibrium quantities
with the equilibrium free energies \cite{jareq97a,jareq97b}. In
\cite{jareq99}, he studied Clausius--Duhem processes via averaging
over the ensemble of microscopic realizations and again obtained
his non-equilibrium equality. As Crooks showed \cite{crooks98},
Jarzynski's equality follows from the assumption that the system
dynamics is Markovian and microscopically reversible. These
conditions are commonly used in computer simulations. In
\cite{crooks99}, Crooks further derived a related fluctuation
theorem of own significance. Quantum counterparts of both the
classical formulations were given by Tasaki \cite{tasaki99}.
Numerous aspects of Jarzynski's equality and some related results
have been addressed in  \cite{ronis07,talhag07,cht11,kafri12}.
Quantum non-equilibrium work relations are still the subject of
active research \cite{cth09,motas10,jareq12,cohen12}.

Concerning the experimental verification of new relations, a
principal difference exists between the classical and quantum
regimes \cite{cht11}. The classical formulations are actually much
easier to validate. Original Jarzynski's equality and Crooks'
theorem has been tested in experiments with individual
biomulecules \cite{liph02,busta05}, macroscopic oscillator
\cite{dcpr05} and electonic system \cite{sytmap12}. The latter was
also studied numerically \cite{pkan13}. The writers of
\cite{npl10} demonstrated feedback manipulations with a Brownian
particle. This allowed to test the corresponding version of
Jarzynski's equality given in \cite{sagu10} (for more details, see
\cite{sagu12}). At the same time, an experimental validation of
the quantum fluctuation relations is still lacking. In
\cite{hskde08,dchfgv13,mdcp13}, some experimental setups to verify
the quantum relations with use of current technology were proposed
(for a discussion, see also sect. VI of \cite{cht11}). Note that
quantum proposals deal with single particles under going unitary
evolution. On the other hand, quantum systems are very sensitive
to noise. To understand the issue properly, it should be
considered from as many aspects as possible. In particular,
theoretical studies of non-equilibrium fluctuations in open
quantum systems may be interesting for possible future
experiments.

Ways to derive Jarzynski's equality are often based on description
within an infinitesimal time scale. In the classical case, we can
use master equation \cite{jareq97b,seif04}, some forms of
deterministic dynamics \cite{jareq00,scholl06} or Markovian
dynamics \cite{crooks98,crooks99}. In the quantum regime, many
approaches were done with somewhat particular assumptions such as
description by Schr\"{o}dinger \cite{jareq12} or master equation
\cite{mukam03}, unitary evolution \cite{tasaki99,motas10}, or the
time-reversal symmetry \cite{cohen12}. An extension of the
Hamiltonian approach to arbitrary quantum systems has been
considered in \cite{cth09}. Notably, in Jarzynski's equality we
actually deal with quantities only at the initial and final
points, without explicit reference to the passage of time. Hence,
we may be interested in obtaining Jarzynski's equality with only
discrete state changes. Certainly, reversible unitary
transformations form a very special class of possible state
changes. The formalism of quantum operations is one of basic tools
in studying dynamics of open quantum systems. This formalism is
especially well adapted to describe discrete state changes. By
such a property, quantum operations are widely used in quantum
information theory \cite{nielsen}. Deterministic processes are
represented by trace-preserving maps known as quantum channels.

The aim of this paper is to study quantum versions of the
Jarzynski equality and related results with use of quantum
operation techniques. The paper is organized as follows. In
Section \ref{sec2}, the preliminary material is reviewed. Basic
notions of quantum operation techniques are recalled. Joint
probability distribution for measurements statistics is discussed
as well. Hence, the averaging rule for considered topics is
obtained. In Section \ref{sec3}, we obtain Jarzynski's equality
under condition that the considered process is represented by a
unital quantum channel. Further, we formulate a fluctuation
theorem of Tasaki--Crooks type for bistochastic map and its
adjoint. We also discuss a heat transfer between two quantum
systems. Initially, the combined system is prepared in the product
state of two particular densities of a special form. We end
Section \ref{sec3} with a short comments about physical relevance
of unital channels and their distinctions from unitary ones. In
Section \ref{sec4}, some equalities with unital channels in the
case of feedback control are obtained. First, we examine the case
of error-free feedback control. Second, we extend the formulation
to a feedback control with classical errors. In Section
\ref{sec5}, we conclude the paper with a summary of results.

\section{Definitions and notation}\label{sec2}

In this section, the required material is presented. First, we
briefly recall basic notions of the formalism of quantum
operations. Second, we describe a general form of the averaging
procedure with a joint probability distribution.

\subsection{States, operators, and quantum channels}\label{ss21}

Let $\lnp(\hh)$ denote the space of linear operators on
finite-dimensional Hilbert space $\hh$. By $\lsa(\hh)$ and
$\lsp(\hh)$, we respectively mean the real space of Hermitian
operators and the set of positive ones. For arbitrary
$\xb,\yb\in\lnp(\hh)$, we define their Hilbert--Schmidt inner
product by \cite{watrous1}
\begin{equation}
\langle\xb{\,},\yb\rangle_{\mathrm{hs}}
:=\Tr(\xb^{\dagger}{\,}\yb)
\ . \label{hsdef}
\end{equation}

We now consider a linear map
$\Phi:{\>}\lnp(\hh_{A})\rightarrow\lnp(\hh_{B})$ that takes
elements of $\lnp(\hh_{A})$ to elements of $\lnp(\hh_{B})$. To
describe a physical process, this map must be completely positive
\cite{nielsen,bengtsson}. Let $\id_{R}$ be the identity map on
$\mathcal{L}(\hh_{R})$, where the space $\hh_{R}$ is assigned to a
reference system. The complete positivity implies that
$\Phi\otimes\id_{R}$ transforms each positive operator into a
positive operator again for each dimension of the extended space.
Any completely positive map can be written in the operator-sum
representation. For all $\xb\in\lnp(\hh_{A})$, we have
\begin{equation}
\Phi(\xb)=\sum\nolimits_{\mu} \kb_{\mu}\xb\kb_{\mu}^{\dagger}
\ . \label{opsm}
\end{equation}
Here, the Kraus operators $\kb_{\mu}$ map the input space
$\hh_{A}$ to the output space $\hh_{B}$. To each linear map
$\Phi:{\>}\lnp(\hh_{A})\rightarrow\lnp(\hh_{B})$, we assign its
adjoint $\Phi^{\dagger}:\lnp(\hh_{B})\rightarrow\lnp(\hh_{A})$ by
the formula \cite{watrous1}
\begin{equation}
\bigl\langle\Phi(\xb),\yb\bigr\rangle_{\mathrm{hs}}=
\bigl\langle\xb{\,},\Phi^{\dagger}(\yb)\bigr\rangle_{\mathrm{hs}}
\ , \label{adjm}
\end{equation}
which holds for all $\xb\in\lnp(\hh_{A})$ and
$\yb\in\lnp(\hh_{B})$. For the linear map (\ref{opsm}), its
adjoint is represented as
\begin{equation}
\Phi^{\dagger}(\yb)=\sum\nolimits_{\mu} \kb_{\mu}^{\dagger}\yb\kb_{\mu}
\ . \label{opsm1}
\end{equation}
The state of an open quantum system is described by a density
matrix $\bro\in\lsp(\hh)$ normalized as $\Tr(\bro)=1$. The input
density matrix $\bro_{A}$ is mapped to the output
$\Phi(\bro_{A})\in\lsp(\hh_{B})$. To be consistent with
probabilistic interpretation, the map $\Phi$ should obey the
condition \cite{nielsen}
\begin{equation}
\sum\nolimits_{\mu} \kb_{\mu}^{\dagger}\kb_{\mu}\leq\pen_{A}
\ , \label{procn}
\end{equation}
where $\pen_{A}$ denotes the identity operator on $\hh_{A}$. By
quantum operations, we mean maps of the form (\ref{opsm}) under
the restriction (\ref{procn}). Deterministic processes are
described by trace-preserving operations, for which the inequality
(\ref{procn}) is saturated and, herewith,
$\Tr\bigl(\Phi(\bro_{A})\bigr)=1$. These maps are usually referred
to as quantum channels \cite{bengtsson}. Most familiar changes of
quantum states are represented by unitary transformations of the
Hilbert space. In this case, the quantum channel has a unique
Kraus operator. To saturate (\ref{procn}), this operator is
inevitably unitary, i.e. $\kb^{\dagger}\kb=\pen$.
Non-trace-preserving quantum operations are also used in quantum
information \cite{nielsen}. To get the output density matrix for
probabilistic operations, we rescale the output as
\begin{equation}
\bro_{B}:=\Tr\bigl(\Phi(\bro_{A})\bigr)^{-1}{\,}\Phi(\bro_{A})
\ . \label{fidm}
\end{equation}
Except for trace-preserving maps, the denominator in (\ref{fidm})
generally depends on the input $\bro_{A}$. On the other hand, the
macroscopic dynamics is deterministic. For these reasons, we
further focus an attention on quantum channels. Two linear maps
$\Phi$ and $\Psi$ can be composed to obtain another linear map
$\Psi\circ\Phi$ such that
\begin{equation}
\bigl(\Psi\circ\Phi\bigr)(\xb):=\Psi\bigl(\Phi(\xb)\bigr)
\ , \label{cmpf}
\end{equation}
for all $\xb\in\lnp(\hh_{A})$. Its operator-sum representation
directly follows from the representations of $\Phi$ and $\Psi$.
The composition of two completely positive maps is completely
positive as well. Hence, such maps form some set with a semigroup
structure. The same observation pertains to quantum channels. An
essential interest to dynamical semigroups was inspired in
studying physical issues such as  master equations, quantum
noises, quantum communication channels, and so on \cite{alfn01}.
Let us consider the operator sum
\begin{equation}
\Phi(\pen_{A})=\sum\nolimits_{\mu} \kb_{\mu}\kb_{\mu}^{\dagger}
\ . \label{uncn}
\end{equation}
Assume that the operator (\ref{uncn}) is a multiple of $\pen_{B}$.
For trace-preserving maps, this condition gives
\begin{equation}
\Phi(\pen_{A})=\frac{d_{A}}{d_{B}}{\>}\pen_{B}
\ . \label{uncn2}
\end{equation}
Here, the integers $d_{A}={\mathrm{dim}}(\hh_{A})$ and
$d_{B}={\mathrm{dim}}(\hh_{B})$ are dimensionalities of $\hh_{A}$
and $\hh_{B}$. When the input and output spaces are of the same
dimensionality, the formula (\ref{uncn2}) gives
$\Phi(\pen_{A})=\pen_{B}$. The latter is usually expresed as the
map $\Phi$ being unital \cite{nielsen}. For the simplest system,
i.e. quantum bit, the depolarizing and phase damping channels are
both unital, whereas the amplitude damping channel is not
\cite{nielsen}. Note that the depolarizing channel, which
represents a decohering qubit, has interesting entropic
characteristics \cite{rastcejp}. Unital trace-preserving maps are
often referred to as bistochastic \cite{bengtsson}. It is easy to
check that the adjoint of bistochastic map is bistochastic as
well. In the following, we derive a useful statement about those
trace-preserving maps that satisfy the condition (\ref{uncn2}).

\subsection{Joint probability distribution and averaging rule}\label{ss22}

The basic idea of statistical physics is to represent a
macroscopic situation of interest by an ensemble of its
microscopic realizations. For a quantum version of Jarzynski's
equality, the corresponding framework was explicitly developed by
Tasaki \cite{tasaki99}. It is convenient to pose an approach
initially for arbitrary observables $\ab\in\lsa(\hh_{A})$ and
$\bab\in\lsa(\hh_{B})$. Their spectral decompositions are
expressed as
\begin{align}
\ab &=\sum\nolimits_{i} a_{i}{\,}|a_{i}\rangle\langle{a}_{i}|
\ , \label{abdc}\\
\bab &=\sum\nolimits_{j} b_{j}{\,}|b_{j}\rangle\langle{b}_{j}|
\ . \label{babdc}
\end{align}
In both the decompositions, the eigenvalues are assumed to be
taken according to their multiplicity. In this regard, we treat
$a_{i}$ and $b_{j}$ as the labels for vectors of the orthonormal
bases $\bigl\{|a_{i}\rangle\bigr\}$ and
$\bigl\{|b_{j}\rangle\bigl\}$. Suppose the system evolution is
represented by quantum channel $\Phi$. If the input state is
described by eigenstate $|a_{i}\rangle$, then the channel output
is $\Phi\bigl(|a_{i}\rangle\langle{a}_{i}|\bigr)$. Then the
probability of being the state $|b_{j}\rangle$ is calculated as
\begin{equation}
p(b_{j}|a_{i})=\langle{b}_{j}|\Phi\bigl(|a_{i}\rangle\langle{a}_{i}|\bigr)|b_{j}\rangle
\ . \label{pcnji}
\end{equation}
This quantity is the conditional probability of the outcome
$b_{j}$ given that the input state was $|a_{i}\rangle$. Due to the
preservation of the trace, we then obtain
\begin{equation}
\sum\nolimits_{j}p(b_{j}|a_{i})=\Tr\Bigl(\Phi\bigl(|a_{i}\rangle\langle{a}_{i}|\bigr)\Bigr)=1
\ . \label{pcnji1}
\end{equation}
Thus, the standard requirement on conditional probabilities is
satisfied with any quantum channel. Further, we suppose that the
input density matrix $\bro_{A}$ has the form
\begin{equation}
\bro_{A}=\sum\nolimits_{i} p(a_{i})|a_{i}\rangle\langle{a}_{i}|
\ , \label{bgf}
\end{equation}
where $\sum_{i}p(a_{i})=1$. The operator (\ref{bgf}) can be
rewritten as a function of the observable $\ab$. Due to the Bayes
rule, one defines the joint probability distribution such that
\begin{equation}
p(a_{i},b_{j})=p(a_{i}){\,}p(b_{j}|a_{i})
\ . \label{pabji}
\end{equation}
This is the probability that we find the system in $i$-th
eigenstate of $\ab$ at the input and in $j$-th eigenstate of
$\bab$ at the output. Consider a function $f(a,b)$ of two
eigenvalues. Extending Tasaki's approach \cite{tasaki99}, we
define an average
\begin{equation}
\bigl\langle{f}(a,b)\bigr\rangle:=
\sum\nolimits_{ij} p(a_{i},b_{j}){\,}f(a_{i},b_{j})
\ . \label{favm}
\end{equation}
Here, angular brackets in the left-hand side signify averaging
over the ensemble of possible pairs of measurement outcomes.
Between these brackets, we will usually omit labels of the
involved variables. In Section \ref{sec4}, however, we will
consider enough complicated protocols. There, the labels will be
all indicated for clearness.

In general, the average (\ref{favm}) does not pertain to
quantum-mechanical expectation values. Namely, the right-hand side
of equation (\ref{favm}) corresponds to specific physical meaning
\cite{tasaki99}. In two simplest cases, however, the average
(\ref{favm}) coincides with the quantum-mechanical expectation
value. Let $a\mapsto{g}(a)$ be some well-defined function. Due to
(\ref{pcnji1}) and (\ref{bgf}), we directly obtain
\begin{equation}
\bigl\langle{g}(a)\bigr\rangle=\sum\nolimits_{i}g(a_{i}){\,}p(a_{i})
=\sum\nolimits_{i}g(a_{i})\langle{a}_{i}|\bro_{A}|a_{i}\rangle=\Tr\bigl(g(\ab){\,}\bro_{A}\bigr)
\ . \label{qexva}
\end{equation}
Using the map linearity and the formulas (\ref{pcnji}) and
(\ref{pabji}), we also write
\begin{equation}
\bigl\langle{g}(b)\bigr\rangle=\sum\nolimits_{j}g(b_{j})\sum\nolimits_{i}p(a_{i})
\langle{b}_{j}|\Phi\bigl(|a_{i}\rangle\langle{a}_{i}|\bigr)|b_{j}\rangle
=\Tr\bigl(g(\bab){\,}\Phi(\bro_{A})\bigr)
\ . \label{qexvb}
\end{equation}

\section{Relations of Jarzynski and Tasaki--Crooks types}\label{sec3}

In this section, we obtain some results connected with Jarzynski's
equality and the Tasaki--Crooks fluctuation theorem. Their
derivation will mainly be based on the condition that the
representing quantum channel is unital. Further, we consider a
heat transfer between two quantum systems.

\subsection{Jarzynski's equality with unital quantum channels}\label{ss31}

Before obtaining Jarzynski's equality, we will formulate a
mathematical result in more abstract form. We consider the
case, in which the input density matrix is expressed as
\begin{equation}
\vrbro_{A}(\alpha):=\Tr\bigl(e^{-\alpha\ab}\bigr)^{-1}e^{-\alpha\ab}
\ . \label{vroin}
\end{equation}
Functional form of such a kind is related to the state of thermal
equilibrium in the Gibbs canonical ensemble. The following exact
relation could be applied beyond the context of Jarzynsky's
equality.

\newtheorem{t31}{Proposition}
\begin{t31}\label{teq31}
Let $\ab\in\lsa(\hh_{A})$, $\bab\in\lsa(\hh_{B})$, and let
$\alpha$ and $\beta$ be real numbers. Suppose the input state is
described by density matrix (\ref{vroin}). If the quantum channel
$\Phi:{\>}\lnp(\hh_{A})\rightarrow\lnp(\hh_{B})$ satisfies the
condition (\ref{uncn2}), then
\begin{equation}
\Bigl\langle
\exp\bigl(\alpha{a}-\beta{b}\bigr)
\Bigr\rangle=\frac{d_{A}}{d_{B}}{\>}
\frac{\Tr\bigl(e^{-\beta\bab}\bigr)}{\Tr\bigl(e^{-\alpha\ab}\bigr)}
\ . \label{prp1}
\end{equation}
\end{t31}

{\bf Proof.} Using the linearity of the map $\Phi$ and the
condition (\ref{uncn2}), we obtain
\begin{equation}
\sum\nolimits_{i} p(b_{j}|a_{i})=
\langle{b}_{j}|\sum\nolimits_{i}\Phi\bigl(|a_{i}\rangle\langle{a}_{i}|\bigr)|b_{j}\rangle
=\langle{b}_{j}|\Phi(\pen_{A})|b_{j}\rangle=\frac{d_{A}}{d_{B}}
\ . \label{sumcp}
\end{equation}
For
$p(a_{i})=\Tr\bigl(e^{-\alpha\ab}\bigr)^{-1}\exp(-\alpha{a}_{i})$
in (\ref{pabji}), we then express the left-hand side of
(\ref{prp1}) as
\begin{equation}
\sum_{ij}\frac{\exp(-\alpha{a}_{i})}{\Tr\bigl(e^{-\alpha\ab}\bigr)}{\>}
p(b_{j}|a_{i}){\,}\exp\bigl(\alpha{a}_{i}-\beta{b}_{j}\bigr)=
\frac{1}{\Tr\bigl(e^{-\alpha\ab}\bigr)}{\>}\sum_{j}\exp(-\beta{b_{j}}){\>}\frac{d_{A}}{d_{B}}
\ . \label{ssmcp}
\end{equation}
The latter term is equal to the right-hand side of (\ref{prp1}).
$\blacksquare$

Due to (\ref{uncn2}), we have evaluated the sum (\ref{sumcp}) in a
closed form. Hence, the claim (\ref{prp1}) was immediately
obtained. Using the result (\ref{prp1}), we will further discuss
the Jarzynski equality. We should emphasize a distinction of the
sum (\ref{sumcp}) from the constraint (\ref{pcnji1}), which takes
the sum with respect to the final events. This point may be
illustrated with the equiprobable distribution
$p(a_{i},b_{j})=(d_{A}d_{B})^{-1}$. Hence, we obtain the
conditional probabilities $p(b_{j}|a_{i})=d_{B}^{-1}$ and the
right-hand side of (\ref{sumcp}). In the context of quantum
channels, such probabilities are realized in the case
$d_{A}\leq{d}_{B}$. Let us consider an isometry
$\vm:{\>}\hh_{A}\rightarrow\hh_{B}$ such that the orthonormal set
$\bigl\{\vm{\,}|a_{i}\rangle\bigr\}$ is mutually unbiased with the
basis $\{|b_{j}\rangle\}$. In other words, one gives
$\bigl|\langle{b}_{j}|\vm|a_{i}\rangle\bigr|=d_{B}^{-1/2}$. The
trace-preserving map $\Phi$ is then defined by (\ref{opsm}) with a
unique Kraus operator $\kb=\vm$. In many cases of interest, the
system dimensionality is not altered during a physical process,
i.e. $d_{A}=d_{B}$.

Let us proceed to an extension of Jarzynsky's equality
\cite{jareq11,jareq97a}. We assume that a thermally contacted
system is acted upon by an external agent. This agent operates
according to a specified protocol. Hence, the Hamiltonian of the
system is time-dependent. The principal system is initially
prepared in the state of thermal equilibrium with a heat
reservoir. Following Tasaki \cite{tasaki99}, we will firstly
assume that the reservoir temperature is also dependent on the
time. The parameters $\beta_{0}$ and $\beta_{1}$ give the inverse
temperature of the reservoir at the initial and final moments,
respectively. Thus, the initial density matrix is
\begin{equation}
\bomg_{0}(\beta_{0})=Z_{0}(\beta_{0})^{-1}e^{-\beta_{0}\hm_{0}}
\ , \label{inidm}
\end{equation}
in terms of the initial Hamiltonian $\hm_{0}$ and the
corresponding partition function
$Z_{0}(\beta_{0})=\Tr\bigl(e^{-\beta_{0}\hm_{0}}\bigr)$. We
further suppose that the transformation of states of the system is
represented by quantum channel $\Phi$ with the same input and
output Hilbert space. In general, the final density matrix
$\Phi\bigl(\bomg_{0}(\beta_{0})\bigr)$ will enough differ from the
matrix
\begin{equation}
\bomg_{1}(\beta_{1})=Z_{1}(\beta_{1})^{-1}e^{-\beta_{1}\hm_{1}}
\ , \label{findm}
\end{equation}
corresponding to equilibrium at the final moment. Here, the
partition function
$Z_{1}(\beta_{1})=\Tr\bigl(e^{-\beta_{1}\hm_{1}}\bigr)$ is
expressed in terms of the final Hamiltonian $\hm_{1}$. By
$\bigl\{\veps_{m}^{(0)}\bigr\}$ and
$\bigl\{\veps_{n}^{(1)}\bigr\}$, we respectively denote
eigenvalues of the Hamiltonians $\hm_{0}$ and $\hm_{1}$. Taking
$d_{A}=d_{B}$ in (\ref{uncn2}) implies that the map is unital. By
obvious substitutions, the formula (\ref{prp1}) then gives
\begin{equation}
\Bigl\langle
{\exp}{\left(\beta_{0}\veps^{(0)}-\beta_{1}\veps^{(1)}\right)}
\Bigr\rangle=\frac{Z_{1}(\beta_{1})}{Z_{0}(\beta_{0})}
\ , \label{tasth}
\end{equation}
provided that the channel $\Phi$ is unital. Assuming unitary
evolution, the relation (\ref{tasth}) has been derived by Tasaki
\cite{tasaki99}. So, we have extended an important formulation to
unital quantum channels.

In the considered context, the term
$w_{nm}=\veps_{n}^{(1)}-\veps_{m}^{(0)}$ is treated as an external
work performed on the principal system during a process. At
constant temperature, i.e. when $\beta_{0}=\beta_{1}=\beta$, we
therefore have
\begin{equation}
\Bigl\langle
\exp\bigl(-\beta{w}\bigr)
\Bigr\rangle=\exp\bigl(-\beta{\,}\Delta{F}\bigr)
\ , \label{jareq0}
\end{equation}
since $F_{t}(\beta)=-\beta^{-1}\ln{Z}_{t}(\beta)$ for $t=0,1$. The
result (\ref{jareq0}) relates, on average, a non-equilibrium
external work with the difference $\Delta{F}=F_{1}-F_{0}$ between
the equilibrium free energies. This formula is the original
Jarzynski equality \cite{jareq97a,jareq97b}. As a consequence, the
basic inequality of thermodynamics can be obtained. Combining
(\ref{jareq0}) with Jensen's inequality for convex function
$x\mapsto\exp(-\beta{x})$ leads to
\begin{equation}
\exp\bigl(-\beta\langle{w}\rangle\bigr)
\leq\exp\bigl(-\beta{\,}\Delta{F}\bigr)
\ . \label{jareq1}
\end{equation}
As the function $x\mapsto\exp(-\beta{x})$ decreases with $x$ for
positive $\beta$, the formula (\ref{jareq1}) gives
$\langle{w}\rangle\geq\Delta{F}$. Thus, total external work
will, on average, exceed the difference between values of the
equilibrium free energy at the final and initial moments. Basing
on Jarzynski's equality, Tasaki also discussed some inequalities
for the von Neumann entropy \cite{tasaki99}.

\subsection{Theorem of Tasaki--Crooks type for bistochastic map and its adjoint}\label{ss32}

Fluctuation theorems are typically used in studying stochastic
processes. They are still the subject of active research
\cite{galc95,lsp99,maes99,lls00}. Such theorems can be
used for deriving information-theoretic results, for instance, Holevo's
bound \cite{kafri12}. For the Tasaki--Crooks fluctuation theorem,
a development with unital quantum channels is reasoned as
follows. For trace-preserving map $\Phi$, its adjoint
$\Phi^{\dagger}$ is unital. To make $\Phi^{\dagger}$
trace-preserving, the map $\Phi$ itself should be unital as well.
For this reason, we focus an attention on bistochastic maps, i.e.
on unital quantum channels. By $\hh$, we denote the Hilbert space
assigned to the principal system. Similarly to (\ref{vroin}), we
introduce the density matrix
\begin{equation}
\vrbro_{B}(\alpha):=\Tr\bigl(e^{-\alpha\bab}\bigr)^{-1}e^{-\alpha\bab}
\ . \label{vrbb}
\end{equation}
Consider two processes obtained by applying the channel $\Phi$ to
the input $\vrbro_{A}$ and the channel $\Phi^{\dagger}$ to the
input $\vrbro_{B}$. In each of the processes, we can ask for a
probability that the difference $(a_{i}-b_{j})$ takes a certain
value. It turns out that the two corresponding probabilities obey
some relation. For a difference between eigenvalues of the
Hamiltonians $\hm_{0}$ and $\hm_{1}$, results of such a kind are
usually referred to as the Tasaki--Crooks theorem
\cite{talhag09,cth09}. It will be convenient, however, to pose a
statement in more abstract form. Let us put the notation. Imposing
the restriction $b_{j}-a_{i}=\Delta$, we fix a difference between
eigenvalues of the observables $\bab$ and $\ab$. By
$P\bigl(b-a=\Delta\big|\Phi,\vrbro_{A}\bigr)$, we denote the
probability of this event given that the channel $\Phi$ represents
an evolution of the input $\vrbro_{A}$. We have the following
result.

\newtheorem{t41}[t31]{Proposition}
\begin{t41}\label{teq41}
Let the quantum channel $\Phi:{\>}\lnp(\hh)\rightarrow\lnp(\hh)$
be unital. For all real $\Delta$, the corresponding probabilities
satisfy
\begin{equation}
e^{-\alpha\Delta}{\,}\Tr\bigl(e^{-\alpha\ab}\bigr){\,}P\bigl(b-a=\Delta\big|\Phi,\vrbro_{A}\bigr)=
\Tr\bigl(e^{-\alpha\bab}\bigr){\,}P\bigl(a-b=-\Delta\big|\Phi^{\dagger},\vrbro_{B}\bigr)
{\>}, \label{pr2eq}
\end{equation}
where the density matrices $\vrbro_{A}$ and $\vrbro_{B}$ are
respectively defined by (\ref{vroin}) and (\ref{vrbb}).
\end{t41}

{\bf Proof.} Assuming action of the channel $\Phi^{\dagger}$, we
use the conditional probability of outcome $a_{i}$ given that the
input state was $|b_{j}\rangle$. Similarly to (\ref{pcnji}), this
probability is written as
\begin{equation}
q(a_{i}|b_{j})=\langle{a}_{i}|\Phi^{\dagger}\bigl(|b_{j}\rangle\langle{b}_{j}|\bigr)|a_{i}\rangle
\ . \label{qcnij}
\end{equation}
The derivation of (\ref{pr2eq}) is based on a simple observation.
Due to the representations (\ref{opsm}) and (\ref{opsm1}), we have
\begin{equation}
\langle{b}|\kb_{\mu}|a\rangle\langle{a}|\kb_{\mu}^{\dagger}|b\rangle=
\langle{a}|\kb_{\mu}^{\dagger}|b\rangle\langle{b}|\kb_{\mu}|a\rangle
\ , \qquad
\langle{b}|\Phi\bigl(|a\rangle\langle{a}|\bigr)|b\rangle=
\langle{a}|\Phi^{\dagger}\bigl(|b\rangle\langle{b}|\bigr)|a\rangle
\ , \label{pqij}
\end{equation}
with arbitrary $|a\rangle,|b\rangle\in\hh$. For all $i$ and $j$,
therefore, one satisfies
\begin{equation}
p(b_{j}|a_{i})=q(a_{i}|b_{j})
\ . \label{pqeq}
\end{equation}
The matrix (\ref{vroin}) has eigenvalues
$p(a_{i})=\Tr\bigl(e^{-\alpha\ab}\bigr)^{-1}\exp(-\alpha{a}_{i})$.
Let $\mcn(\Delta)$ be the set of ordered pairs $(j,i)$ such that
$b_{j}-a_{i}=\Delta$. By (\ref{pqeq}), the left-hand side of
(\ref{pr2eq}) is then rewritten as
\begin{align}
\sum_{(j,i)\in\mcn(\Delta)}\exp(-\alpha{b}_{j}+\alpha{a}_{i}){\,}\Tr\bigl(e^{-\alpha\ab}\bigr){\,}p(a_{i}){\,}p(b_{j}|a_{i})&=
\sum_{(j,i)\in\mcn(\Delta)}\exp(-\alpha{b}_{j}){\,}p(b_{j}|a_{i})
\nonumber\\
&=\Tr\bigl(e^{-\alpha\bab}\bigr)\sum_{(j,i)\in\mcn(\Delta)}q(b_{j}){\,}q(a_{i}|b_{j})
{\>}. \label{ppqq}
\end{align}
Here the numbers
$q(b_{j})=\Tr\bigl(e^{-\alpha\bab}\bigr)^{-1}\exp(-\alpha{b}_{j})$
are eigenvalues of $\vrbro_{B}$. For all the ordered pairs
$(j,i)\in\mcn(\Delta)$, we have $a_{i}-b_{j}=-\Delta$. Combining
this with the definition (\ref{qcnij}), the right-hand
side of (\ref{ppqq}) is equal to the right-hand side of
(\ref{pr2eq}). $\blacksquare$

We can also write some relation with the completely mixed state
$\bro_{*}=\pen/d$, where $d={\rm{dim}}(\hh)$ and
$p(a_{i})=q(b_{j})=1/d$. Repeating the above reasons, we obtain
\begin{equation}
P\bigl(b-a=\Delta\big|\Phi,\bro_{*}\bigr)=P\bigl(a-b=-\Delta\big|\Phi^{\dagger},\bro_{*}\bigr)
\ . \label{pr2cm}
\end{equation}
It seems that the matrices of the form (\ref{vroin}) and the
$\bro_{*}$ are the only two forms, for
which a closed relation between the two probabilities could be
written.

The statement of Proposition \ref{teq41} immediately leads to a
theorem of Tasaki--Crooks type. Let us consider a thermally
insulated system acted upon by an external field. As above, the
operators $\hm_{0}$ and $\hm_{1}$ are respectively the initial and
final Hamiltonians. For a unital quantum channel $\Phi$, we apply
this channel itself to the equilibrium state (\ref{inidm}) and the
adjoint $\Phi^{\dagger}$ to the equilibrium state (\ref{findm}).
Differences $w_{nm}=\veps_{n}^{(1)}-\veps_{m}^{(0)}$ are treated
as possible values of an external work performed on the system
during the former process. By obvious substitution, the formula
(\ref{pr2eq}) gives
\begin{equation}
e^{-\beta{w}}{\,}Z_{0}(\beta){\,}P\bigl(\veps^{(1)}-\veps^{(0)}=w|\Phi,\bomg_{0}\bigr)=
Z_{1}(\beta){\,}P\bigl(\veps^{(0)}-\veps^{(1)}=-w|\Phi^{\dagger},\bomg_{1}\bigr)
\ , \label{pr23}
\end{equation}
with the inverse temperature $\beta$ of heat reservoir. Using
$\Delta{F}=F_{1}-F_{0}$, the relation (\ref{pr23}) can be
rewritten in the form
\begin{equation}
\frac{P\bigl(\veps^{(1)}-\veps^{(0)}=
w|\Phi,\bomg_{0}\bigr)}{P\bigl(\veps^{(0)}-\veps^{(1)}=-w|\Phi^{\dagger},\bomg_{1}\bigr)}
=\exp\bigl(\beta{w}-\beta\Delta{F}\bigr)
\ , \label{pr24}
\end{equation}
when probabilities are non-zero for taken $w$. If the channel
$\Phi$ represents a unitary evolution, then its adjoint
$\Phi^{\dagger}$ represents the inverse unitary evolution. In the
case of unitary transformations, the exact relation (\ref{pr23})
with the two probabilities has been derived by Tasaki
\cite{tasaki99}. It is a quantum analog of previous Crooks'
formulation \cite{crooks98}. In the literature, the above
statement is often referred to as the Tasaki--Crooks fluctuation
theorem \cite{talhag07,cth09}. Thus, we have obtained an extension
of the Tasaki--Crooks fluctuation theorem to unital quantum
channels. Finally, we emphasize that use of adjoint maps in the
formulation inevitably leads to unital channels. Indeed, the
adjoint map is trace-preserving only for a bistochastic map.

\subsection{Notes on heat transfer between two systems}\label{ss33}

We now apply the above results to a heat transfer between two
quantum systems. The following analysis is an extension of related
results of the paper \cite{tasaki99}. It is convenient, however,
to put a derivation in a more abstract manner. Consider two
systems with the Hilbert spaces $\hh_{A}$ and $\hh_{B}$,
respectively. So, we write the initial density matrix of the
combined system as
\begin{equation}
\vrbro_{AB}:=\vrbro_{A}(\alpha)\otimes\vrbro_{B}(\beta)
=\Tr\bigl(e^{-\alpha\ab}\bigr)^{-1}{\,}\Tr\bigl(e^{-\beta\bab}\bigr)^{-1}
{\,}e^{-\alpha\ab}\otimes{e}^{-\beta\bab}
\ , \label{abvr}
\end{equation}
with $\ab\in\lsa(\hh_{A})$ and $\bab\in\lsa(\hh_{B})$. Here, we
used the matrices (\ref{vroin}) and (\ref{vrbb}), but the latter
with $\beta$ instead of $\alpha$. We also assume that a state
change of the combined system is represented by the channel
$\Psi$, with the input space $\hh_{AB}=\hh_{A}\otimes\hh_{B}$ and
the output space $\hh_{C}$. Let us take quantum channels that
satisfy condition of the form (\ref{uncn2}), namely
\begin{equation}
\Psi(\pen_{AB})=\frac{d_{A}d_{B}}{d_{C}}{\>}\pen_{C}
\ , \label{abcn2}
\end{equation}
where $\pen_{AB}=\pen_{A}\otimes\pen_{B}$ and $\pen_{C}$ are the
corresponding identities. The operator
$\alpha\ab\otimes\pen_{B}+\pen_{A}\otimes\beta\bab$ is Hermitian
for real $\alpha$ and $\beta$. It has eigenvalues
$\alpha{a}_{i}+\beta{b}_{j}$ and eigenstates
$|a_{i}b_{j}\rangle=|a_{i}\rangle\otimes|b_{j}\rangle$. In terms
of this operator, the matrix (\ref{abvr}) reads
\begin{equation}
\vrbro_{AB}:=\Tr\Bigl(\exp\bigl(-\alpha\ab\otimes\pen_{B}-\pen_{A}\otimes\beta\bab\bigr)\Bigr)^{-1}
\exp\bigl(-\alpha\ab\otimes\pen_{B}-\pen_{A}\otimes\beta\bab\bigr)
\ , \label{abvr1}
\end{equation}
since the summands $\alpha\ab\otimes\pen_{B}$ and
$\pen_{A}\otimes\beta\bab$ commute. For arbitrary
$\cab\in\lsa(\hh_{C})$, the following conclusion can be written as
a variety of (\ref{prp1}). If the quantum channel
$\Psi:{\>}\lnp(\hh_{AB})\rightarrow\lnp(\hh_{C})$ satisfies
(\ref{abcn2}) and the input is given by (\ref{abvr}), then
\begin{equation}
\Bigl\langle
\exp\bigl(\alpha{a}+\beta{b}-c\bigr)
\Bigr\rangle=\frac{d_{A}d_{B}}{d_{C}}{\>}
\frac{\Tr\bigl(e^{-\cab}\bigr)}{\Tr\bigl(e^{-\alpha\ab}\bigr){\,}\Tr\bigl(e^{-\beta\bab}\bigr)}
\ . \label{dabc}
\end{equation}
In the case $\hh_{C}=\hh_{AB}$ and
$\cab=\alpha\ab\otimes\pen_{B}+\pen_{A}\otimes\beta\bab$, the
formula (\ref{dabc}) becomes
\begin{equation}
\Bigl\langle
{\exp}{\left(\alpha(a-a^{\prime})+\beta(b-b^{\prime})\right)}
\Bigr\rangle=1
\ , \label{dabc1}
\end{equation}
where the set
$\bigl\{\alpha{a}_{i}^{\prime}+\beta{b}_{j}^{\prime}\bigr\}$
denotes the spectrum of $\cab$. The condition (\ref{abcn2}) is
reduced here to $\Psi(\pen_{AB})=\pen_{AB}$.

We now consider the following question. Each of two separated
systems is initially prepared in equilibrium at the inverse
temperatures $\beta_{0}$ and $\beta_{1}$, respectively. Their
density matrices are therefore written as (\ref{inidm}) and
(\ref{findm}), whence the product
$\bomg_{0}(\beta_{0})\otimes\bomg_{1}(\beta_{1})$ is the input
total state. Suppose that the systems further interact via unital
quantum channel $\Psi$. By obvious substitutions into
(\ref{dabc1}), for the described process we obtain
\begin{equation}
\Bigl\langle
{\exp}{\left(\beta_{0}(\veps^{(0)}-\veps^{\prime(0)})+\beta_{1}(\veps^{(1)}-\veps^{\prime(1)})\right)}
\Bigr\rangle=1
\ . \label{dabc2}
\end{equation}
Assuming unitary evolution, this result has been given in
\cite{tasaki99}. Thus, we have extended the previous result to
unital quantum channels. In the paper \cite{tasaki99}, the
relation (\ref{dabc2}) is presented with a certain physical
interpretation. Let us consider the quantity
\begin{equation}
\Delta{S}:=\Bigl\langle
\beta_{0}(\veps^{\prime(0)}-\veps^{(0)})+\beta_{1}(\veps^{\prime(1)}-\veps^{(1)})\Bigr\rangle
\ . \label{desdf}
\end{equation}
The terms $\bigl\langle\veps^{\prime(0)}-\veps^{(0)}\bigr\rangle$
and $\bigl\langle\veps^{\prime(1)}-\veps^{(1)}\bigr\rangle$ give,
on average, the change of self-energy of the corresponding
systems. Suppose that changes in the inverse temperatures of the
systems are sufficiently small and a contribution of interaction
energy to the total entropy is negligible. In such a situation,
the quantiy $\Delta{S}$ estimates an averaged change of the total
entropy \cite{tasaki99}. Combining (\ref{dabc2}) with Jensen's
inequality for convex function $x\mapsto\exp(-x)$, we obtain
\begin{equation}
\exp\bigl(-\Delta{S}\bigr)\leq1
\ , \qquad
\Delta{S}\geq0
\ . \label{des0}
\end{equation}
Thus, we have arrived at the well-known inequality of
thermodynamics. An extension of Clausius' inequality to arbitrary
non-equilibrium processes beyond linear response was obtained in
\cite{de10}. In the paper \cite{tasaki99}, Tasaki also discussed a
way for constructing more detailed bounds on $\Delta{S}$ from
below. We only emphasize here that this can be achieved with unital
quantum channels.

Thus, we have shown that Jarzynki's equality and many related
results remain valid in the case, when the evolution of a quantum
system is represented by unital quantum channels. In this regard,
we do not need in the unitarity assumption, which is much more
restrictive. Let us discuss briefly a significance of unital
channels and their distinctions from unitary ones. Unitary
channels form a very special class of quantum channels. If the
given quantum channel is invertible for all inputs, then it is
unitary with necessity \cite{preskill}. Here, we do not mean the
invertibility for the prescribed input, as treated in
the data processing inequality \cite{nielsen}. With respect to the
map composition, the set of all unitary channels has a structure
of group. To consider dynamics of open quantum systems, we have to
left out invertibility. Assuming the most general form of state
changes, we should focus an attention on dynamical semigroups.
Indeed, the composition of two trace-preserving maps is also
trace-preserving. It is not insignificant that similar observation
pertains to unitality. Namely, the composition of two unital maps
is unital as well. Thus, bistochastic maps form the set with a
semigroup structure. For all such maps, the Jarzynski equality and
the Tasaki--Crooks fluctuation theorem are still valid. The
simplest examples of unital and non-unitary channels are the phase
damping and depolarizing qubit channels. Both are examples of
subtle and important quantum processes. For example, key effects in the
Schr\"{o}dinger cat-atom system may be modeled as phase damping
\cite{nielsen}. The depolarizing channel represents a typical case
of decohering qubit \cite{preskill}. Another reason to consider
unital channels was pointed out in connection with the
Tasaki--Crooks fluctuation theorem. The role of unitality in
statistical physics of small quantum systems deserves further
investigations.

\section{Jarzynski--Sagawa--Ueda relations with unital quantum channels}\label{sec4}

In this section, we develop some of the above results in the case,
when the agent makes measurement followed by a feedback. First,
error-free feedback control is considered. Second, we analyze the
case, in which classical errors occur in the measurement process.

\subsection{Error-free feedback control}\label{ss41}

We assume that the agent performs a quantum measurement and
further acts according to the measurement outcome.
For classical systems, this topic has been considered by Sagawa
and Ueda \cite{sagu10,sagu12}. For quantum systems, relations of
such a kind were examined by Morikuni and Tasaki \cite{motas10}.
Let us recall required material on quantum measurements. In
general, the quantum measurement is posed as a set
$\{\nm_{\mu}\}$ of measurement operators, acting on the space of
the measured system \cite{nielsen}. If the pre-measurement state
is described by $\bro$, then the probability of $\mu$-th outcome
is $\Tr\bigl(\nm_{\mu}^{\dagger}\nm_{\mu}\bro\bigr)$. The
corresponding post-measurement state is described by density
matrix
\begin{equation}
\bro_{\mu}^{\prime}=
\Tr\bigl(\nm_{\mu}^{\dagger}\nm_{\mu}\bro\bigr)^{-1}\nm_{\mu}\bro\nm_{\mu}^{\dagger}
\ . \label{post}
\end{equation}
Note that number of measurement outcomes can arbitrarily exceed
dimensionality of the Hilbert space. This possibility is crucial
for many quantum protocols \cite{nielsen}. The set of measurement
operators satisfies the completeness relation
\begin{equation}
\sum\nolimits_{\mu}\nm_{\mu}^{\dagger}\nm_{\mu}=\pen
\ . \label{cplr}
\end{equation}
When the measurement operators are mutually orthogonal projectors,
the above scheme obviously leads to traditional projective
measurements. We will also say about quantum measurements that
fulfill the condition
\begin{equation}
\sum\nolimits_{\mu}\nm_{\mu}\nm_{\mu}^{\dagger}=\pen
\ . \label{pclr}
\end{equation}
The projective measurements all satisfy the condition
(\ref{pclr}). Moreover, for Hermitian measurement operators this
condition coincides with the completeness relation (\ref{cplr}).

Let us pose formally the protocol with error-free feedback
control. For the sake of simplicity, we will assume the same input
and output space for all adopted operations. Further, we use an
appropriate number of observables $\bab_{\mu}\in\lsa(\hh)$, each
with its eigenbasis $\bigl\{|b_{j}^{(\mu)}\rangle\bigr\}$. We also
define the corresponding density matrices
$\vrbro_{B}^{(\mu)}(\alpha)$ by the formula (\ref{vrbb}) with
$\bab_{\mu}$ instead of $\bab$. We shall consider the following
procedure.

\begin{itemize}
\item[(i)]{At the first stage, the agent applies quantum channel $\Phi $ to the input $\vrbro_{A}$ given by (\ref{vroin}).}
\item[(ii)]{At the second stage, one performs the quantum measurement on the output $\Phi(\vrbro_{A})$. For $\mu$-th outcome, its probability is $p(\mu)={\Tr}{\left(\nm_{\mu}^{\dagger}\nm_{\mu}\Phi(\vrbro_{A})\right)}$ and the post-measurement state is $p(\mu)^{-1}\nm_{\mu}\Phi(\vrbro_{A})\nm_{\mu}^{\dagger}$.}
\item[(iii)]{At the third stage, the agent applies the prescribed quantum channel $\Psi_{\mu}$ to $\mu$-th post-measurement state given that $\mu$-th outcome occurs.}
\item[(iv)]{At the fourth stage, one measures the observable $\bab_{\mu}$ on the third-stage output $p(\mu)^{-1}\Psi_{\mu}{\left(\nm_{\mu}\Phi(\vrbro_{A})\nm_{\mu}^{\dagger}\right)}$. With this pre-measurement state, the outcome $b_{j}^{(\mu)}$ is obtained with the probability
\begin{equation}
p(\mu)^{-1}\langle{b}_{j}^{(\mu)}|\Psi_{\mu}{\left(\nm_{\mu}\Phi(\vrbro_{A})\nm_{\mu}^{\dagger}\right)}|b_{j}^{(\mu)}\rangle
\ . \label{prvr}
\end{equation}}
\end{itemize}

Multiplying (\ref{prvr}) by $p(\mu)$, i.e. by $\mu$-th outcome
probability, we obtain the probability of the outcome
$b_{j}^{(\mu)}$ for the input (\ref{vroin}). The latter
probability can be represented as the sum
\begin{equation}
\sum\nolimits_{i}p(a_{i},b_{j}^{(\mu)})=\sum\nolimits_{i}p(a_{i}){\,}p({b}_{j}^{(\mu)}|a_{i})
\ . \label{pamb}
\end{equation}
Here, we introduce the conditional probability
\begin{equation}
p({b}_{j}^{(\mu)}|a_{i})=
\langle{b}_{j}^{(\mu)}|\Psi_{\mu}{\left(\nm_{\mu}\Phi(|a_{i}\rangle\langle{a}_{i}|)\nm_{\mu}^{\dagger}\right)}|b_{j}^{(\mu)}\rangle
\ , \label{pbac}
\end{equation}
given that the input was $|a_{i}\rangle$. These probabilities
satisfy the required condition
\begin{equation}
\sum\nolimits_{j\mu}p({b}_{j}^{(\mu)}|a_{i})=
\sum\nolimits_{\mu}{\Tr}{\left(\nm_{\mu}^{\dagger}\nm_{\mu}\Phi(|a_{i}\rangle\langle{a}_{i}|)\right)}=1
\ . \label{pbac1}
\end{equation}
We have used the preservation and the cyclic property of the trace
and the completeness relation (\ref{cplr}). For clarity, all the
labels of involved variables will be explicitly indicated between
the angular brackets.

\newtheorem{t51}[t31]{Proposition}
\begin{t51}\label{teq51}
Let the above protocol be applied to the input (\ref{vroin}), and
let the quantum channel $\Phi:{\>}\lnp(\hh)\rightarrow\lnp(\hh)$
be unital. For arbitrary quantum measurement $\{\nm_{\mu}\}$ and
quantum channels $\Psi_{\mu}$, we have
\begin{equation}
\Bigl\langle
\frac{\Tr\bigl(e^{-\alpha\ab}\bigr)}{\Tr\bigl(e^{-\alpha\bab_{\mu}}\bigr)}
{\>}\exp\bigl(\alpha{a}_{i}-\alpha{b}_{j}^{(\mu)}\bigr)
\Bigr\rangle=
\sum\nolimits_{\mu}\Tr\Bigl(\vrbro_{B}^{(\mu)}(\alpha){\,}\Psi_{\mu}\bigl(\nm_{\mu}\nm_{\mu}^{\dagger}\bigr)\Bigr)
\ . \label{efrfc}
\end{equation}
\end{t51}

{\bf Proof.} Since the channel $\Phi$ is unital and
$\langle{b}_{j}^{(\mu)}|\exp\bigl(-\alpha{b}_{j}^{(\mu)}\bigr)=\langle{b}_{j}^{(\mu)}|\exp(-\alpha\bab_{\mu})$,
we first write
\begin{align}
\frac{\Tr\bigl(e^{-\alpha\ab}\bigr)}{\Tr\bigl(e^{-\alpha\bab_{\mu}}\bigr)}
{\,}\sum_{i} p(a_{i}){\,}p({b}_{j}^{(\mu)}|a_{i}){\,}\exp\bigl(\alpha{a}_{i}-\alpha{b}_{j}^{(\mu)}\bigr)
&=\frac{\exp\bigl(-\alpha{b}_{j}^{(\mu)}\bigr)}{\Tr\bigl(e^{-\alpha\bab_{\mu}}\bigr)}
{\,}\langle{b}_{j}^{(\mu)}|\Psi_{\mu}{\left(\nm_{\mu}\Phi(\pen)\nm_{\mu}^{\dagger}\right)}|b_{j}^{(\mu)}\rangle
\nonumber\\
&=\langle{b}_{j}^{(\mu)}|\vrbro_{B}^{(\mu)}(\alpha){\,}\Psi_{\mu}\bigl(\nm_{\mu}\nm_{\mu}^{\dagger}\bigr)|b_{j}^{(\mu)}\rangle
\ . \label{cism}
\end{align}
Summing the right-hand side of (\ref{cism}) with respect to $j$,
we get
$\Tr\Bigl(\vrbro_{B}^{(\mu)}(\alpha){\,}\Psi_{\mu}\bigl(\nm_{\mu}\nm_{\mu}^{\dagger}\bigr)\Bigr)$.
The latter leads to (\ref{efrfc}) after further summing with
respect to $\mu$. $\blacksquare$

The statement of Proposition \ref{teq51} is written in a general
form. We now apply this result to a thermally insulated system,
which is operated by the agent with feedback control. For each of
the possible ways, we introduce the corresponding Hamiltonian
$\hm_{\mu}$ and equilibrium state
\begin{equation}
\bomg_{\mu}(\beta)=Z_{\mu}(\beta)^{-1}e^{-\beta\hm_{\mu}}
\ , \qquad
Z_{\mu}(\beta)=\Tr\bigl(e^{-\beta\hm_{\mu}}\bigr)
\ . \label{bmgm}
\end{equation}
The protocol is applied to the input $\bomg_{0}(\beta)$, whereas
the averaging is taken over the final states $\bomg_{\mu}(\beta)$
with $\mu\neq0$. Let us introduce the associated work
$w_{nm}^{(\mu)}=\veps_{n}^{(\mu)}-\veps_{m}^{(0)}$. By obvious
substitution, we rewrite the equality (\ref{efrfc}) as
\begin{equation}
\Bigl\langle
{\exp}{\left(-\beta{w}_{nm}^{(\mu)}+\beta(F_{\mu}-F_{0})\right)}
\Bigr\rangle=\gamma
\ , \label{efrfc1}
\end{equation}
where $F_{\mu}(\beta)=-\beta^{-1}\ln{Z}_{\mu}(\beta)$ and the
parameter $\gamma$ is defined by
\begin{equation}
\gamma:=\sum\nolimits_{\mu}\Tr\Bigl(\bomg_{\mu}(\beta){\,}\Psi_{\mu}\bigl(\nm_{\mu}\nm_{\mu}^{\dagger}\bigr)\Bigr)
{\,}. \label{gamdf}
\end{equation}
The formula (\ref{efrfc1}) is a quantum counterpart of one of the
results originally formulated in \cite{sagu10}. With unitary
transformations of quantum states, the relation (\ref{efrfc1}) was
derived in \cite{motas10}. For two projective measurements
and any trace-preserving map between them, a similar relation was
considered in \cite{kafri12}. Thus, we have extended an important
non-equilibrium equality to the case, when the unital channel
$\Phi$ acts at the stage (i) and arbitrary channels $\Psi_{\mu}$
act at the stage (iii). Let us discuss a particular case without
any feedback. Here, the same channel $\Psi$ is applied for all the
outcomes and final states are always compared with
$\bomg_{1}(\beta)$. If the channel $\Psi$ is unital and the POVM
$\{\nm_{\mu}\}$ obeys (\ref{pclr}), then we have
$\gamma=\Tr\bigl(\bomg_{1}(\beta)\bigr)=1$. It is Jarzynski's
equality (\ref{jareq0}) with an intermediate quantum measurement.

\subsection{Feedback control with classical errors}\label{ss42}

The above result can be modified to the case, when measurement
outcomes are registered with some randomness. By probabilities
$r(\nu|\mu)$, we represent purely classical nature of errors at
the stage (ii). The quantity $r(\nu|\mu)$ is the conditional
probability of mis-interpretation of actual $\mu$-th outcome as
registered $\nu$-th one. Recall the concept of mutual information.
The pointwise mutual information is defined as \cite{fano61}
\begin{equation}
I_{\mu\nu}:=\ln\frac{p(\mu,\nu)}{p(\mu)p(\nu)}=\ln\frac{r(\mu|\nu)}{p(\mu)}=\ln\frac{r(\nu|\mu)}{p(\nu)}
\ , \label{pwi0}
\end{equation}
where we used Bayes' rule. This quantity can take positive or
negative values, vanishing for $p(\mu,\nu)=p(\mu)p(\nu)$.
Averaging (\ref{pwi0}) with the joint probability distribution, we
obtain the mutual information
\begin{equation}
\langle{I}_{\mu\nu}\rangle=\sum_{\mu\nu}p(\mu,\nu){\,}\ln\frac{p(\mu,\nu)}{p(\mu)p(\nu)}
\ . \label{mi0}
\end{equation}
The mutual information is extensively treated
in information theory \cite{fano61}.

In the case of feedback control with classical errors, we should
average with the joint probability distribution
\begin{equation}
p(a_{i},\mu,b_{j}^{(\nu)})=p(a_{i}){\,}p(\mu,b_{j}^{(\nu)}|a_{i})
\ , \label{pamb1}
\end{equation}
in which the latter conditional probability is written as
\begin{equation}
p(\mu,b_{j}^{(\nu)}|a_{i})=r(\nu|\mu){\,}
\langle{b}_{j}^{(\nu)}|\Psi_{\nu}{\left(\nm_{\mu}\Phi(|a_{i}\rangle\langle{a}_{i}|)\nm_{\mu}^{\dagger}\right)}|b_{j}^{(\nu)}\rangle
\ . \label{pbac2}
\end{equation}
For error-free feedback control, we have
$r(\nu|\mu)=\delta_{\nu\mu}$. Here, the right-hand side of
(\ref{pbac2}) is non-zero only for $\nu=\mu$, when it is reduced
to the right-hand side of (\ref{pbac}). Similarly to
(\ref{pbac1}), by means of $\sum_{\nu}r(\nu|\mu)=1$ we also obtain
\begin{equation}
\sum\nolimits_{j\mu\nu}p(\mu,b_{j}^{(\nu)}|a_{i})= 1
\ . \label{pbac11}
\end{equation}

\newtheorem{t61}[t31]{Proposition}
\begin{t61}\label{teq61}
Let the above protocol be applied to the input (\ref{vroin}), and
let classical errors at the stage (ii) be represented by the
conditional probability $r(\nu|\mu)$. If the quantum channel
$\Phi:{\>}\lnp(\hh)\rightarrow\lnp(\hh)$ is unital, then
\begin{equation}
\Bigl\langle
\frac{\Tr\bigl(e^{-\alpha\ab}\bigr)}{\Tr\bigl(e^{-\alpha\bab_{\nu}}\bigr)}
{\>}\exp\bigl(\alpha{a}_{i}-\alpha{b}_{j}^{(\nu)}\bigr)
\Bigr\rangle=
\sum\nolimits_{\mu\nu}r(\nu|\mu){\,}\Tr\Bigl(\vrbro_{B}^{(\nu)}(\alpha){\,}\Psi_{\nu}\bigl(\nm_{\mu}\nm_{\mu}^{\dagger}\bigr)\Bigr)
{\,}. \label{rfc1}
\end{equation}
If the quantum channels
$\Psi_{\mu}:{\>}\lnp(\hh)\rightarrow\lnp(\hh)$ are also all unital
and the POVM $\{\nm_{\mu}\}$ obeys (\ref{pclr}) then
\begin{equation}
\Bigl\langle
\frac{\Tr\bigl(e^{-\alpha\ab}\bigr)}{\Tr\bigl(e^{-\alpha\bab_{\nu}}\bigr)}
{\>}\exp\bigl(\alpha{a}_{i}-\alpha{b}_{j}^{(\nu)}-I_{\nu\mu}\bigr)
\Bigr\rangle=1
\ . \label{rfc2}
\end{equation}
\end{t61}

{\bf Proof.} Similarly to the proof of Proposition \ref{teq51}, we
first obtain the relation
\begin{equation}
\frac{\Tr\bigl(e^{-\alpha\ab}\bigr)}{\Tr\bigl(e^{-\alpha\bab_{\nu}}\bigr)}
{\,}\sum_{ij} p(a_{i}){\,}p(\mu,b_{j}^{(\nu)}|a_{i}){\,}\exp\bigl(\alpha{a}_{i}-\alpha{b}_{j}^{(\nu)}\bigr)
=r(\nu|\mu){\,}\Tr\Bigl(\vrbro_{B}^{(\nu)}(\alpha){\,}\Psi_{\nu}\bigl(\nm_{\mu}\nm_{\mu}^{\dagger}\bigr)\Bigr)
{\,}, \label{cism1}
\end{equation}
provided that the channel $\Phi$ is unital. After summing with
respect to $\mu$ and $\nu$, we get the first claim (\ref{rfc1}).
Multiplying (\ref{cism1}) by $\exp(-I_{\nu\mu})=p(\nu)/r(\nu|\mu)$
and further summing with respect to $\mu$, one yields
\begin{equation}
\frac{\Tr\bigl(e^{-\alpha\ab}\bigr)}{\Tr\bigl(e^{-\alpha\bab_{\nu}}\bigr)}
{\,}\sum_{ij\mu} p(a_{i}){\,}p(\mu,b_{j}^{(\nu)}|a_{i}){\,}\exp\bigl(\alpha{a}_{i}-\alpha{b}_{j}^{(\nu)}-I_{\nu\mu}\bigr)=
p(\nu){\,}\Tr\Bigl(\vrbro_{B}^{(\nu)}(\alpha){\,}\Psi_{\nu}(\pen)\Bigr)
{\,} , \label{cism2}
\end{equation}
under the condition (\ref{pclr}). If $\Psi_{\nu}(\pen)=\pen$ for
all $\nu$, summing of (\ref{cism2}) with respect to $\nu$ finally
gives $\sum_{\nu}p(\nu)=1$. $\blacksquare$

It is essential for (\ref{rfc1}) and (\ref{rfc2}) that
corresponding quantum operations be unital. The proof of the
result (\ref{rfc1}) assumes this property only for the channel at
the first stage (i). The proof of the result (\ref{rfc2}) has also
assumed this property for the measurement at the stage (ii) and
for all the channels at the stage (iii). Let us proceed to a
thermally insulated system, which is operated by the agent with
feedback control. In the considered case, the formula (\ref{rfc1})
gives
\begin{equation}
\Bigl\langle
{\exp}{\left(-\beta{w}_{nm}^{(\nu)}+\beta(F_{\nu}-F_{0})\right)}
\Bigr\rangle=\widetilde{\gamma}
\ , \label{efrfc2}
\end{equation}
where the parameter $\widetilde{\gamma}$ is written as
\begin{equation}
\widetilde{\gamma}:=\sum\nolimits_{\mu\nu}r(\nu|\mu){\,}
\Tr\Bigl(\bomg_{\nu}(\beta){\,}\Psi_{\nu}\bigl(\nm_{\mu}\nm_{\mu}^{\dagger}\bigr)\Bigr)
{\,}. \label{gamdf2}
\end{equation}
Replacing $\gamma$ with $\widetilde{\gamma}$, the result
(\ref{efrfc2}) is completely similar to (\ref{efrfc1}). For
error-free feedback control, the term $\widetilde{\gamma}$ becomes
$\gamma$ due to $r(\nu|\mu)=\delta_{\nu\mu}$. The formulas
(\ref{gamdf}) and (\ref{gamdf2}) express the parameters in a
closed form. Notably, these parameters are experimentally
measurable quantities. The original Sagawa--Ueda formulation with
feedback control is classical. It has been tested experimentally
\cite{npl10}. The quantum results (\ref{efrfc1}) and
(\ref{efrfc2}) may be used in the context of future experiments.
They hold when the channel $\Phi$ is unital. If the quantum
operations at the stages (ii) and (iii) are unital as well, then
\begin{equation}
\Bigl\langle
{\exp}{\left(-\beta{w}_{nm}^{(\nu)}+\beta(F_{\nu}-F_{0})-I_{\nu\mu}\right)}
\Bigr\rangle=1
\ . \label{efrfc11}
\end{equation}
This is a quantum counterpart of one of the results obtained by
Sagawa and Ueda \cite{sagu10}. In the paper \cite{motas10}, the
relation (\ref{efrfc11}) has been presented within a unitary
evolution under assumption that the measurement obeys
(\ref{pclr}). The formula (\ref{efrfc2}) has also been given in
\cite{motas10} with unitary transformations of the states. Thus,
we have extended quantum Jarzynski--Sagawa--Ueda relations to
unital quantum channels.

\section{Conclusions}\label{sec5}

We have considered the Jarzynski equality and related fluctuation
results from the viewpoint of quantum operation techniques. The
formalism of quantum operations was developed to describe dynamics
of open quantum systems. Since study of open systems is one of key
aims of statistical physics, the language of quantum operations
may offer new insights. In the paper, we have considered some
advances of the mentioned viewpoint. It is very essential that
representing quantum channels be unital. Assuming this property,
the Jarzynski equality is still valid. A fluctuation theorem of
Tasaki--Crooks type has been formulated for bistochastic map and
its adjoint. With unital quantum channels, we also apply the
formalism to the problem of heat transfer between two quantum
systems. Some equalities with unital channels have also been
derived in the case of feedback control. Error-free feedback and
feedback with classical errors are both considered. Hence, quantum
Jarzynski--Sagawa--Ueda relations have been generalized to unital
quantum channels. The obtained expressions may be useful in
experimental tests. Thus, the formalism of quantum operations
provides a suitable framework for studying non-equilibrium
relations in open quantum systems. The described approach is
interesting in own rights as well as from the viewpoint of future
experimental validation of the quantum non-equilibrium results.
First, novel aspects of the problem could be analyzed in terms of
quantum stochastic maps. In particular, this approach is well
adapted to describe noise in open quantum systems. Indeed, for
existing or future proposals, we should estimate a degree of
environmental noise and its influence on experimental results.
Second, the formalism of quantum operations is now common
language to represent state transformations in quantum information
processing. In principle, practical achievements in quickly
growing area of quantum information may be used in experimental
tests of non-equilibrium relations in quantum systems.

\acknowledgments
The present author is grateful to anonymous referee for
constructive criticism.

\medskip

{\it{Note added.}} After the present paper was submitted I learned
about the recent work \cite{albash}, in which the significance of
unitality is emphasized as well. In the work \cite{albash}, the
authors formulate a general fluctuation theorem and further show
that some previous results follow from this theorem. My
formulations and derivation methods are different from and, in
certain respects, complementary to those given in \cite{albash}.
The authors of \cite{albash} also describe results of a related
experiment with superconducting flux qubits.

\end{document}